\newif\ifpp
\pptrue % or
\ifpp
\documentclass[11pt,a4paper]{article}
\else
\documentclass{svjour3}                     % onecolumn (standard format)
\fi
\pdfoutput=1
\usepackage{graphicx}
\usepackage{siunitx}
\usepackage{tabularx,booktabs} 
\usepackage{authblk}
\usepackage{hyperref}

\newcommand{\dd}{\mathrm{d}}

\newcommand{\ua}{\uparrow}
\newcommand{\da}{\downarrow}

\ifpp
\fi
\newcommand{\avrg}[1]{\left<  #1 \right>}
\title{Extraction of Azimuthal Asymmetries using Optimal Observables
\footnote{accepted for publication in EPJC}}

\ifpp
\author[1,2,3]{J\"org Pretz}
\author[1,2]{Fabian M\"uller}
\affil[1]{Institut f\"ur Kernphysik, Forschungszentrum J\"ulich, 52425 J\"ulich, Germany}
\affil[2]{III. Physikalisches Institut B, RWTH Aachen University, 52056 Aachen, Germany}
\affil[3]{JARA-FAME (Forces and Matter Experiments), Forschungszentrum J\"ulich and RWTH Aachen University}
\fi

\begin{document}

\ifpp
\else
\author{J\"org Pretz  \and Fabian M\"uller}
\institute{J\"org Pretz \at
            Institut f\"ur Kernphysik, Forschungszentrum J\"ulich, 52425 J\"ulich, Germany \\
        III. Physikalisches Institut B, RWTH Aachen University, 52056 Aachen, Germany \\
            JARA-FAME, Forschungszentrum J\"ulich und RWTH Aachen University \\
         \email{pretz@physik.rwth-aachen.de}  \and
            Fabian M\"uller \at
           Institut f\"ur Kernphysik, Forschungszentrum J\"ulich, 52425 J\"ulich, Germany \\
            III. Physikalisches Institut B, RWTH Aachen University, 52056 Aachen, Germany           
}
\fi

\date{}

\maketitle

\begin{abstract}
  Azimuthal asymmetries play an important role in scattering processes with polarized particles.
  This paper introduces a new procedure using event weighting to extract these asymmetries.
  It is shown that the resulting estimator has several advantages in terms of statistical accuracy, bias, 
  assumptions on acceptance and luminosities
  compared to other estimators discussed in the literature.

\ifpp
\bigskip \noindent
{\bf keywords:}
data analysis, polarisation, likelihood, parameter estimation, event weighting, analyzing power,  minimal variance bound, 
Cram\'er-Rao bound, optimal observables, generalized method of moments
\else

\fi

\end{abstract}

\section{Introduction and motivation}
This paper describes the extraction of an azimuthal asymmetry $\epsilon$ occurring in an event distribution
given by
\begin{equation}\label{eq:N}
  N(\vartheta,\varphi)  = \frac{1}{2\pi} \, \mathcal{L} \, a(\vartheta,\varphi) \, \sigma_0(\vartheta) \, (1 + \epsilon(\vartheta) \cos(\varphi)) \, .
\end{equation}
The variables in eq.~\ref{eq:N} are defined in table~\ref{tab:sigma}.
Event distributions of this type appear for example in scattering processes of a transversally polarised beam
on a spin 0 target~\cite{Ohlsen:1972zz}.
\begin{table}[hpb]
\begin{center}
  \begin{tabular}{|l|l|}
    \hline
  variable & meaning \\
  \hline
    $N(\vartheta,\varphi)$ & number of events observed \\
  $\avrg{N(\vartheta,\varphi)}$ & expectation value of number of events \\
%    $\sigma(\vartheta,\varphi)$ & cross section \\
     $\sigma_0(\vartheta)$ & unpolarized cross section \\
    $\vartheta$ & polar angle \\
  $\varphi$ & azimuthal angle, $\varphi=0$ corresponds to positive $x$-direction \\    
   $\epsilon = PA$ & asymmetry parameter to be determined \\
    $P$  & beam polarization \\
    $A(\vartheta)$  & analyzing power \\
    $\mathcal{L}$ & luminosity \\
    $a(\vartheta,\varphi)$ & acceptance \\
\hline
  \end{tabular}
\caption{Definitions of variables used in eq.~\ref{eq:N}.\label{tab:sigma}}
\end{center}
\end{table}
The parameter $\epsilon$ is the product of the polarisation and an analyzing power, $\epsilon = P A$.
Once $\epsilon$ is determined one can either determine the polarisation $P$ if the analyzing power $A$ is known,
or vice versa.
To cancel systematic effects, one usually takes two data sets with opposite polarisations, e.g. polarisation up ($P^\ua$) 
and down ($P^\da$).
The acceptance factor $a(\vartheta, \varphi)$ may have an arbitrary dependence on the $\varphi$ and $\vartheta$.
The only assumption is that the acceptance is the same for the two data sets.

In this paper a new estimator using event weights and a $\chi^2$-minimization is introduced.
  The method is an application of optimal observables discussed in refs.~\cite{Atwood:1991ka,Tkachov:2014kya},
but it also takes into account luminosity and acceptance effects.
The paper is organized as follows.
In section~\ref{sec:methods} several estimators to determine $\epsilon$ (i.e. $P$ or $A$) are discussed
and compared.
Section~\ref{sec:w} introduces the new method.
%In Sec.~\ref{sec:mc} the results obtained are verified with MC simulations.
Possible extensions of this new weighting/fitting method are discussed in Sec.~\ref{sec:ext}.

\section{Estimators to determine azimuthal asymmetries}\label{sec:methods}
In general one can distinguish two classes of  estimators: estimators using event counts, discussed in 
subsection~\ref{sec:counts} 
and estimators using event weights, discussed in subsection~\ref{sec:w}.

\subsection{Estimators using event counts}\label{sec:counts}
\begin{figure}
\includegraphics[width=\textwidth]{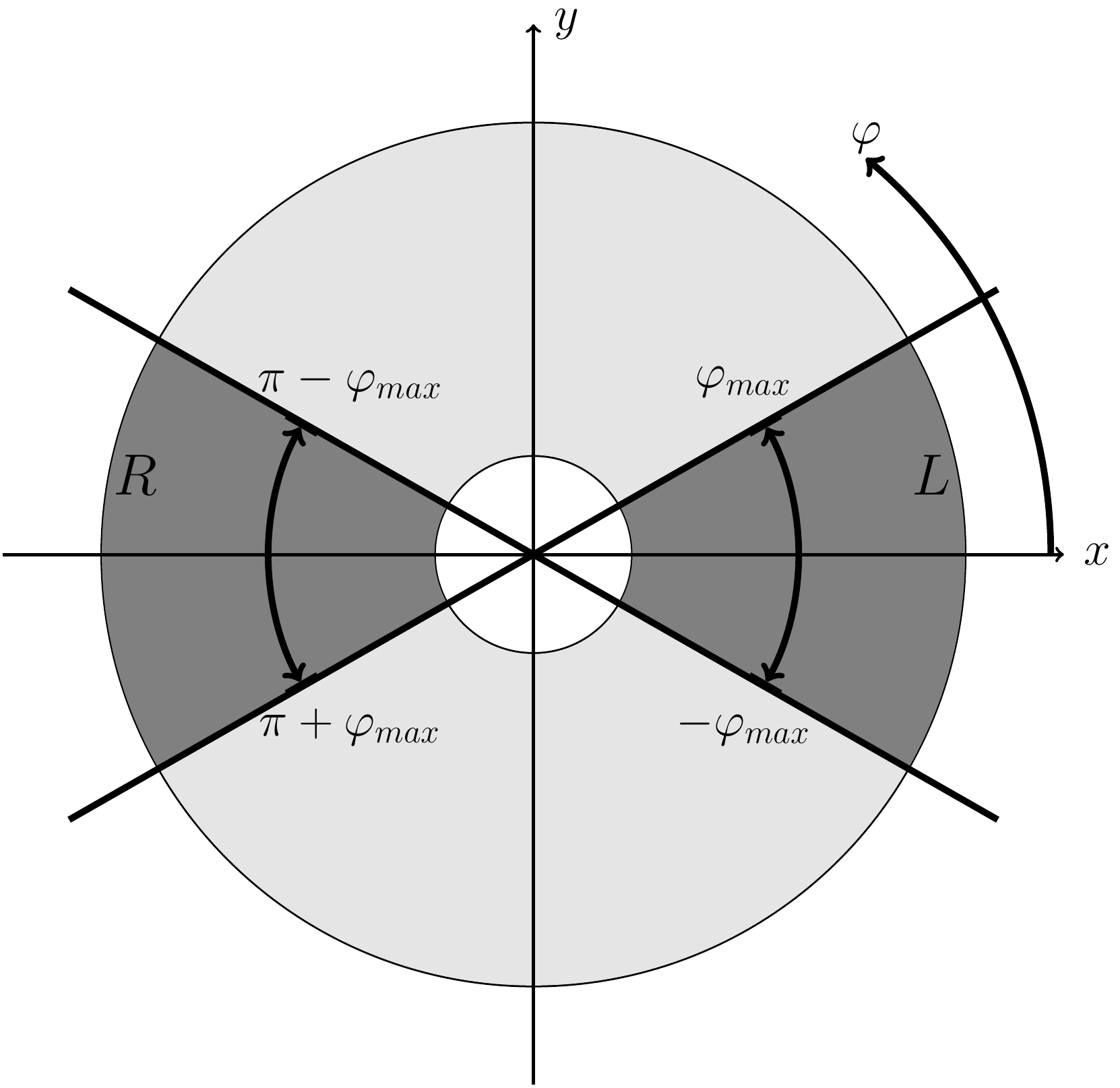}
\caption{Definition of azimuthal angle and accepted events.
  The beam moves in $z$-direction, i.e. out of the plane.\label{fig:phi}}
\end{figure}

Here events around $\varphi=0$ and $\varphi=\pi$ as indicated by the dark region in figure~\ref{fig:phi} enter 
the analysis.
The expectation value for the number of events in the left ($L$) part of the detector is given by:
\begin{eqnarray}
  \avrg{N_L^\ua} &=& \frac{1}{2\pi} \, \int_{-\varphi_{max}}^{\varphi_{max}} \mathcal{L}^{\ua}  a(\varphi)  \sigma_0 \left(1 + P^\ua A \cos(\varphi) \right) \dd \varphi \\
%  &=&  \mathcal{L}^{\ua}  \sigma_0 \left( \int_{-\varphi_{max}}^{\varphi_{max}} a(\varphi) \dd \varphi
%  + P^\ua A \int_{-\varphi_{max}}^{\varphi_{max}} a(\varphi) \cos(\varphi)  \dd \varphi \right)\\
  &=&  \mathcal{L}^{\ua}  a_L \sigma_0  
  \left( 1 +   \avrg{\cos(\varphi)}_L  P^\ua A \right)
\end{eqnarray}
with
\begin{eqnarray*}
  a_L &=& \frac{1}{2\pi} \, \int_{-\varphi_{max}}^{\varphi_{max}} a(\varphi)   \dd \varphi  \quad \mbox{and} \\
   \avrg{\cos(\varphi)}_L  &=& \frac{\int_{-\varphi_{max}}^{\varphi_{max}} a(\varphi) \cos(\varphi)  \dd \varphi}{\int_{-\varphi_{max}}^{\varphi_{max}} a(\varphi)   \dd \varphi} \, .
\end{eqnarray*}  
To simplify the notation the $\vartheta$-dependence is dropped.
Similar equations exist for  $\avrg{N_R^\ua}$, $\avrg{N_L^\da}$ and $\avrg{N_R^\da}$.

In the cross ratio 
\begin{equation}\label{eq:cr}
 \delta = \frac{\avrg{N^{\ua}_L}  \avrg{N^{\da}_R}}{\avrg{N^{\ua}_R}  \avrg{N^{\da}_L}} = \frac{(1 +  \avrg{\cos(\varphi)}_L P^{\ua} A) (1 + \avrg{\cos(\varphi)}_R P^{\da} A)}{(1 +  \avrg{\cos(\varphi)}_R P^{\ua} A) (1 +  \avrg{\cos(\varphi)}_L P^{\da} A)} \, 
\end{equation}
 introduced in ref.~\cite{Ohlsen:1973wf},
 the usually unknown luminosities, acceptances and unpolarized cross section cancel.
Replacing the expectation values 
 by the actual measured event counts, the following estimator for the analyzing power $A$ can be derived
 \begin{eqnarray}\
%          \hat{A} = \frac{P^\da (C_R - C_L \delta)- P^\ua (C_L - C_R \delta)-
%           \sqrt{ (P^\da(-C_R + C_L \delta) P^\ua (- C_L + C_R \delta))^2 
%            + 4 (\delta-1) C_L C_R  P^\da P^\ua}}
%              {2 C_L C_R P^\da P^\ua (\delta-1)   }
%
%
%   \hat{A} = \frac{P^\da (C_R  - C_L \delta)+  P^\ua (C_L - C_R \delta) - 
%   \sqrt{-4 C_L C_R  P^\da P^\ua (\delta-1)^2  + (P^\da (C_R-C_L \delta) + P^\ua (C_L-C_R \delta))^2}}{2 C_L C_R P^\da P^\ua (\delta-1)}
%
   \hat{A} &=& \frac{X-\sqrt{X^2- 2 Y (\delta-1)}}{Y} \, , \quad \mbox{with}   \label{eq:A_cr} \\
      && X =  P^\da \left( \avrg{\cos(\varphi)}_R  - \avrg{\cos(\varphi)}_L \delta \right)+  P^\ua \left(\avrg{\cos(\varphi)}_L - \avrg{\cos(\varphi)}_R \delta \right)  \quad \mbox{and} \nonumber \\
   && Y = 2 \avrg{\cos(\varphi)}_L \avrg{\cos(\varphi)}_R P^\da P^\ua(\delta-1) \, . \nonumber
   %
% mathematica: A -> ((Pdn (cr  - cl delta)+  Pup (cl - cr delta) - 
%   Sqrt[-4 cl cr (delta-1)^2 Pdn Pup + (Pdn (cr-cl*delta) + Pup (cl-cr delta))^2])/(2 cl cr Pdn Pup (delta-1)))
 \end{eqnarray}
% with
% \[
%  C_L =  \avrg{\cos(\varphi)}_L    \, , \quad C_R=  \avrg{\cos(\varphi)}_R
% \]
Note that to evaluate $\avrg{\cos(\varphi)}_{L,R}$ information on the acceptance is needed.
This method was for example applied in ref.~\cite{Bagdasarian:2014mdj}.
Here bins of $\Delta \varphi=\pm 30$ degrees  were used.

 Another possibility is to consider estimators of the type 
\begin{equation}\label{eq:A_linear}
  \hat{A} = \frac{1}{P \avrg{\cos(\varphi)}} \, \frac{N_L^{\ua(\da)}- N_R^{\ua(\da)}}{N_L^{\ua(\da)} + N_R^{\ua(\da)}}
\quad \quad
 \mbox{or}
\quad \quad
  \hat{A} = \frac{1}{P \avrg{\cos(\varphi) }} \, \frac{N_{L(R)}^\ua- N_{L(R)}^\da}{N_{L(R)}^\ua + N_{R(L)}^\da}
\end{equation}
where various corrections have to be applied in order to compensate for acceptance and luminosity
difference between the two data sets. These type of estimators were used in refs.~\cite{Czyzykiewicz:2006km,Sakamoto:1996xdz}.

Common to these estimators is that they reach the same statistical error $\sigma$.
In general it is more convenient to work with the figure of merit (FOM) defined by $\mbox{FOM} = \sigma^{-2}$.
To evaluate the FOM we make a few assumptions 
to simplify the notation:
First, $P^\ua = -P^\da$, in addition we assume that one takes roughly the same number of events
in both polarisation configurations. We also assume a uniform acceptance in $\varphi$. 
It is straight forward to derive formulas dropping these assumptions but the expressions are getting cumbersome.
These assumptions do not change the overall conclusions comparing different estimators.
Instead of discussing the FOM on $A$, we will discuss the FOM of $\epsilon$.

Error propagation from eqs.~\ref{eq:A_cr} or \ref{eq:A_linear} leads to  
\begin{equation}\label{eq:fom_cr_exact}
  \mbox{FOM}_{\epsilon}^{\mathrm{counts}} = N_{\mathrm tot} \, \frac{\avrg{\cos(\varphi)}^2}{1-\avrg{\cos(\varphi)}^2 \epsilon^2} \, 
\end{equation}
where $N_{\mathrm tot}$ is the total number of events entering the analysis. Details of the calculation are given in 
app.~\ref{app:fom_cnt}.
Neglecting the term with $\epsilon$, one finds:
\begin{eqnarray}
  \mbox{FOM}^{\mathrm{counts}}_\epsilon &=&  N_{\mathrm tot} \, \avrg{\cos(\varphi)}^2  \\
   &=& N_0  \, \frac{2 \varphi_{max}}{ \pi}  \, 
    \left(  \frac{ \int_{-\varphi_{max}}^{\varphi_{max}}  \cos(\varphi)\dd \varphi}{  \int_{-\varphi_{max}}^{\varphi_{max}}  \dd \varphi } \right)^2
    = N_0 \,      \frac{2 \sin^2(\varphi_{max})}{\pi \varphi_{max}}  \label{eq:fom_cr} 
%      & \stackrel{\varphi_{max}=\pi/2}{=} &N_0  \, \frac{4}{\pi^2}  \label{eq:fom_cr1}
\end{eqnarray}
where $N_0 = \int_0^{2 \pi} a \sigma_0 (\mathcal{L}^\ua + \mathcal{L}^\da) \dd \varphi$ is the total 
number of events available in both polarisation states.
Thus $N_{\mathrm tot} = N_0 (2 \varphi_{max})/(\pi)$ is the total number of events entering the analysis.

\begin{figure}
 \includegraphics[width=\textwidth]{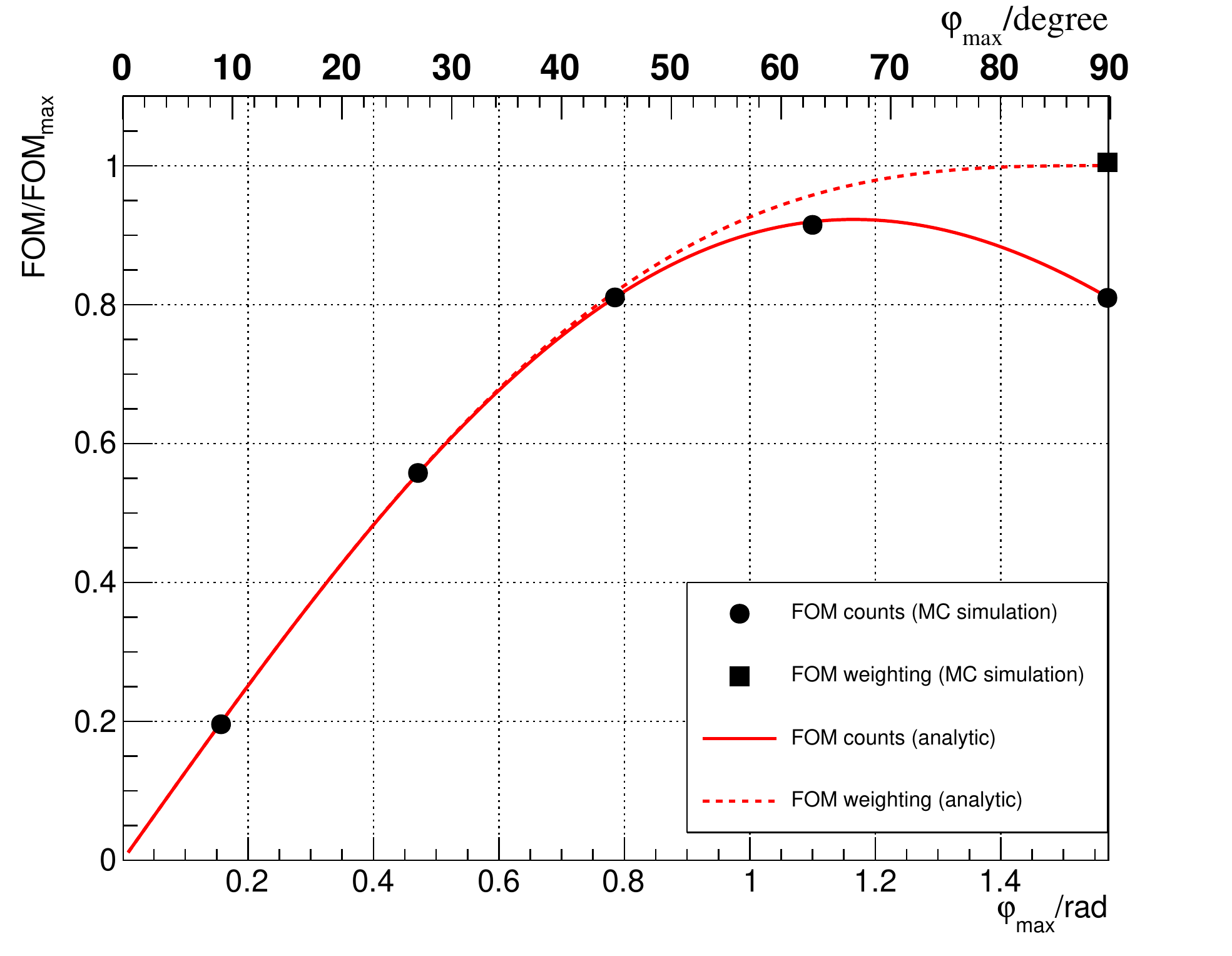}
 \caption{Figure of merit (FOM) for estimators using event counts and event weighting,
   calculated analytically (lines) and from MC simulation (symbols).\label{fig:fom}}
\end{figure}

The full line in figure~\ref{fig:fom} shows the FOM calculated according to eq.~\ref{eq:fom_cr} 
for different $\varphi$-ranges.
%assuming constant acceptance in $\varphi$. 
Increasing $\varphi_{max}$, the FOM increases first. Around $\varphi_{max}\approx 65$ degrees it starts to decrease.
The reason is that one adds more and more events  where $\cos(\varphi)$ is small. These events
carry less information on $\epsilon$ and dilute the sample in the way the analysis is performed.
This clearly shows that this cannot be the optimal strategy.
In the next section estimators will be discussed where the FOM reaches the dashed line,
which corresponds to the Cram\'er-Rao bound.

\subsection{Estimators using event weights} \label{sec:w}
In this section estimators are discussed which use event weights
instead of event counts as in the previous subsection.
In ref.~\cite{Besset:1979sh} weighted sums $\sum_i \cos(\varphi_i)$ are introduced in order to extract $\epsilon$.
To cancel acceptance effects the authors propose to combine the event distributions from the two polarisation states.
They do not address the question how to deal with different luminosities in the two different polarisation states.
The method was applied in ref.~\cite{Daum:2001kk} where an azimuthal symmetry of the detector is assumed.
It is also shown in  ref.~\cite{Besset:1979sh} that with this weighting procedure the FOM reaches  the Cram\'er-Rao 
bound as does the unbinned likelihood method. An unbinned likelihood method was used in ref.~\cite{Adolph:2012nw}.
It is not straight forward to apply because the probability density function is not completely known.
Acceptance effects have to be verified using a Monte Carlo simulation.

Now a new method, reaching the Cram\'er-Rao bound as well, is introduced.
The advantage is that no knowledge about the acceptance is required (as long as it is the same for both data sets, 
as in any other method)
and no corrections concerning the luminosities have to be applied. On the contrary,
information on the acceptance and luminosity factor $\mathcal{L} \sigma_0 a_0$
are obtained in parallel to $\epsilon$ in this method.

We consider the following six observables  
\begin{eqnarray*}
   \sum_{i=1}^{N^\ua} \cos^n(\varphi_i) \quad  \mbox{and}  \quad  \sum_{i=1}^{N^\da} \cos^n(\varphi_i) \quad , 
   \mbox{with}  \quad n=0,1,2 \, .
\end{eqnarray*}
The sums run over the number of events in the given polarisation state including all azimuthal angels.
Note that $n=0$ corresponds just to the number of events observed,
$n=1 (2)$ are higher moments and correspond to the sum over events weighted with $\cos(\varphi) (\cos^2(\varphi))$.

For an arbitrary acceptance in $\varphi$ 
we can write the following Fourier series:
\begin{equation}\label{eq:acc}
  a(\varphi) = a_0 + \sum_{n=1}^{\infty} a_n \cos(n \varphi) + b_n \sin(n \varphi) \, .
\end{equation}

The expectation values of these observables are given by
\footnotesize
\begin{eqnarray}
  \avrg{N^\ua} &=&  \frac{1}{2\pi} \, \mathcal{L^\ua}\sigma_0 \int_0^{2 \pi}
  \left[ a_0 + \sum_{i=n}^{\infty} a_n \cos(n \varphi) + b_n \sin(n \varphi)  \right] \left(1 + P^\ua A \cos(\varphi) \right) \dd\varphi  \nonumber\\
  &=&  \mathcal{L^\ua} \sigma_0 a_0 \left(1  + \frac{a_1}{2 a_0} P^\ua A \right) \, ,   \label{eq:nup}  \\
  && \nonumber \\
  && \nonumber \\    
  {\avrg{\sum_\ua \cos(\varphi_i)}} &=&  \frac{1}{2\pi} \, \mathcal{L^\ua} \sigma_0\int_0^{2 \pi}
  \cos(\varphi) \left[ a_0 + \sum_{n=1}^{\infty} a_n \cos(n \varphi) + b_n \sin(n \varphi)  \right] \left(1 + P^\ua A \cos(\varphi_i) \right) \dd\varphi\nonumber \\
   &=&  \frac{1}{2} \, \mathcal{L^\ua} \sigma_0  a_0 \left(P^\ua A  \left(1  + \frac{a_2}{2 a_0} \right) + \frac{a_1}{a_0}  \right) \, ,\\
  && \nonumber \\
  && \nonumber \\
  {\avrg{\sum_\ua \cos^2(\varphi_i)}} &=&  \frac{1}{2\pi} \, \mathcal{L^\ua} \sigma_0 \int_0^{2 \pi}
  \cos^2(\varphi) \left[ a_0 + \sum_{n=1}^{\infty} a_n \cos(n \varphi) + b_n \sin(n \varphi)  \right] \left(1 + P^\ua A \cos(\varphi) \right) \dd\varphi\nonumber  \\
  &=&  \frac{1}{2} \, \mathcal{L^\ua}  \, \sigma_0 a_0 \left(  \left(1+ \frac{a_2}{2a_0} \right) + \frac{1}{4}  \, \frac{3 a_1+a_3}{a_0}  P^\ua A \right) \, .  \label{eq:nup2}
  \end{eqnarray}
\normalsize
Similar expressions hold for the expectation values $ \avrg{N^\da}, \avrg{\sum_\da \cos(\varphi_i)}, \avrg{\sum_\da \cos^2(\varphi_i)}$ of the second polarisation state by replacing $P^\ua$
with $P^\da$. The integrals extend over all azimuthal angles from 0 to 2$\pi$. It is also possible to
apply the method for a limited range as in the previous section. In this case the integrals would extend
over $[-\varphi,\varphi]$ and $[\pi-\varphi,\pi+\varphi]$ (dark region in figure~\ref{fig:phi}).

Assuming that the polarisations  $P^\ua$ and  $P^\da$ are known,
using a $\chi^2$ minimization comparing the expectation values with the observables, one 
can determine the following 6 unknown parameters:
\[
  (\mathcal{L}^\ua \sigma_0 a_0), (\mathcal{L}^\da \sigma_0 a_0), \frac{a_1}{a_0}, \frac{a_2}{a_0}, \frac{a_3}{a_0}, A \, .
\]
The $\chi^2$ is given by:
\begin{eqnarray}
    \chi^2 &=& (\vec y_{\mathrm obs} - \vec y_{\mathrm model}) \, C^{-1} \, (\vec y_{\mathrm obs} - \vec y_{\mathrm model})^T \label{eq:chi2}
  \end{eqnarray}
with 
\begin{eqnarray}
   \vec{y}_{\mathrm obs} &=& \left( N^\ua, \sum_{\ua} \cos(\varphi_i),\sum_{\ua} \cos^2(\varphi_i),
   N^\da, \sum_{\da} \cos(\varphi_i),\sum_{\da} \cos^2(\varphi_i) \right) \, , \nonumber \\
  \vec{y}_{\mathrm model} &=& \left( {\avrg{N^\ua}},{\avrg{\sum_{\ua} \cos(\varphi_i)}},{\avrg{\sum_\ua \cos^2(\varphi_i)}},
     {\avrg{N^\da}},{\avrg{\sum_\da \cos(\varphi_i)}},{\avrg{\sum_\da \cos^2(\varphi_i)}} \right)   \, . \nonumber
\end{eqnarray}  
The covariance matrix $C$ of the observables is given  in app.~\ref{app:cov}.
The easiest way to obtain values for the parameters is to minimize eq.~\ref{eq:chi2} numerically
although analytic, but cumbersome, expressions exist for the parameters.
The numerical solution is also preferred in view of possible extensions of the method discussed in Sec.~\ref{sec:ext}, where analytic solutions may not exist.

The FOM, calculated using the same conditions as used for $\mbox{FOM}^{\mathrm{counts}}_\epsilon$ in eq.~\ref{eq:fom_cr},
is derived in app.~\ref{app:fom_w}. The final result is:
\begin{equation}\label{eq:fom_w_exact}
\mbox{FOM}^{\mathrm{weighting}}_\epsilon = N_{\mathrm tot} \frac{\avrg{\cos^2(\varphi)}^2} {\avrg{\cos^2(\varphi)} - \avrg{\cos^4(\varphi)} \epsilon^2} \, .
\end{equation}
Neglecting the term with $\epsilon$ one finds:
\begin{eqnarray}\label{eq:fom_w}
  \mbox{FOM}^{\mathrm{weighting}}_\epsilon &=&  N_{\mathrm tot} \avrg{\cos^2{\varphi}} \\
   &=& N_0  \, \frac{ 2 \varphi_{max}}{\pi}   \,  \frac{\int_{-\varphi_{max}}^{\varphi_{max}} \cos^2(\varphi) \dd \varphi}{  \int_{-\varphi_{max}}^{\varphi_{max}}  \dd \varphi }
= N_0  \, \frac{\varphi_{max} + \sin(\varphi_{max}) \cos(\varphi_{max})}{ \pi}  \, .
%      & \stackrel{\varphi_{max}=\pi/2}{=} &  \frac{1}{2} N_0  \, .
\end{eqnarray}
It is shown as a dashed line in figure~\ref{fig:fom}.
At small $\varphi_{max}$ the FOM of counting and weighting estimators coincide, at larger $\varphi_{max}$,
$\mbox{FOM}^{\mathrm{weighting}}_\epsilon$ keeps increasing.

%\subsection{Summary of Methods}

\subsection{General discussion on the figure of merit}\label{sec:fom}
%We start this section with a more general discussion on the statistical accuracy $\sigma_{\epsilon}$.
%Formulas are derived for $\epsilon = PA$. 
In this subsection we make some general remarks about the FOM reachable for
event distributions of the type
\begin{equation}\label{eq:n}
   n(\varphi) = \alpha(\varphi) \left(1 \pm \beta(\varphi) \epsilon  \right) \, .
\end{equation}
As shown in ref.~\cite{Pretz:2011qb} the estimator
\begin{equation}\label{eq:epshat}
   \hat{\epsilon} = \frac{\sum_\ua w(\varphi_i) - \sum_\da w(\varphi_i)}{\sum_\ua w(\varphi_i) \beta(\varphi_i) + \sum_\da w(\varphi_i) \beta(\varphi_i)}
\end{equation}
is bias free, where $w(\varphi)$ is an arbitrary weight function.
The FOM  is given by
\[
   \mbox{FOM}^{w}_{\epsilon} = N_{\mathrm tot} \frac{\avrg{w \beta}^2}{\avrg{w^2(1-\epsilon^2 \beta^2)}} \, .
\]
The choice $w=1$, or to be more precise $w=1$ if the event enters the analysis 
and $w=0$ else, results in
\begin{equation}\label{eq:fom_cnt_gen}
   \mbox{FOM}_{\epsilon}^{w=1} = N_{\mathrm tot} \frac{\avrg{\beta}^2}{\avrg{(1-\epsilon^2 \beta^2)}} \, .
\end{equation}
The choice $w=\beta$ leads to the largest FOM (in the limit $\epsilon \ll 1$) reaching the  Cram\'er-Rao bound:
\begin{equation}\label{eq:fom_w_gen}
   \mbox{FOM}_{\epsilon}^{w=\beta} = N_{\mathrm tot} \frac{\avrg{\beta^2}^2}{\avrg{\beta^2(1-\epsilon^2 \beta^2)}} \, .
\end{equation}
Translated to azimuthal asymmetries the factor $\beta(\varphi)$ equals $\cos(\varphi)$.
The two FOMs given in eq.~\ref{eq:fom_cr_exact}, sec.~\ref{sec:counts} and eq.~~\ref{eq:fom_w_exact}, sec.~\ref{sec:w}
are identical to the FOMs
of eqs.~\ref{eq:fom_cnt_gen} and \ref{eq:fom_w_gen}, respectively.

%%%%xxxxxxxxxxxxxx

\subsection{Results of simulations}\label{sec:mc}
In this subsection we crosscheck the results of the previous subsections and discuss possible bias
with the help of Monte Carlo simulations.
A Monte Carlo simulation with  $10^6$ events in total was performed by generating data according to eq.~\ref{eq:N}
for two polarizations states with $P^\ua=0.5$ and $P^\da=-0.5$ and $A=0.2$. The acceptance was once assumed to be uniform
in $\varphi$ and once the following parameterization
\begin{eqnarray}
  a(\varphi) = 1 &+&  0.3 \cos(\varphi) \, \,-0.2 \sin(\varphi) \nonumber \\
                 &-&  0.3 \cos(2\varphi) + 0.1 \sin(2\varphi) \nonumber \\
                 &+&  0.2 \cos(3\varphi) + 0.2 \sin(3\varphi) \nonumber \\
                 &-&  0.1 \cos(4\varphi) + 0.1 \sin(4\varphi)  \label{eq:acc1} \, 
\end{eqnarray}
was used. In the analysis it is assumed that $a(\varphi)$ is unknown.
Table~\ref{tab:res_mc} summarizes the results found using a MINUIT minimization in ROOT~\cite{Brun:1997pa}
to minimize $\chi^2$ in eq.~\ref{eq:chi2}.
One sees that with the weighting/fitting method, one recovers the input analyzing power
and the acceptance factors. No bias is observed.
The cross ratio method, using events in the range $-1.2 < \varphi_{max} <1.2$ to maximize the FOM (see figure~\ref{fig:fom}),
gives an unbiased result for $A$ only in the case of uniform $\varphi$ acceptance as expected,
since $\avrg{\cos(\varphi)}$ was calculated under this assumption.

The circles in figure~\ref{fig:fom} show the FOM obtained from the RMS of 1000 simulations where the analyzing 
powers was calculated according to eq.~\ref{eq:A_cr} for various values of $\varphi_{max}$.
The square symbol is the FOM obtained from MINUIT using the weighting/fitting procedure.
There is perfect agreement between the simulations and analytic formulas.

\begin{table}
  \begin{center}
\begin{tabular}{|l|r|
    S[                 table-number-alignment = right,
                 separate-uncertainty = true,
                 table-figures-uncertainty = 5,
                 table-figures-decimal = 5
    ]|
    S[                 table-number-alignment = right,
                 separate-uncertainty = true,
                 table-figures-uncertainty = 5,
                 table-figures-decimal = 5
]|    
  }
  \hline
parameter &  input value  & \text{cross ratio, counting}                  & \text{weighting/fit}    \\
\hline
\multicolumn{4}{|c|}{uniform acceptance}  \\  
\hline
$A$  &  0.2      &   0.2030 \pm 0.0029    &  0.2030   \pm 0.0028          \\
$a_1/a_0$ &  0  &   &  -0.0002 \pm 0.0014                 \\ 
$a_2/a_0$  & 0  &    &  -0.0002 \pm 0.0014 \\
$a_3/a_0$  & 0  &    &  -0.0002 \pm 0.0028 \\
\hline
\multicolumn{4}{|c|}{non-uniform acceptance, eq.~\ref{eq:acc1}}  \\   
 \hline
 $A$  &  0.2      &   0.1910 \pm 0.0031    &  0.2036 \pm 0.0031   \\
$a_1/a_0$ &  0.3  &   &  0.3003 \pm 0.0013                 \\ 
$a_2/a_0$  & $-0.3$  &    &  -0.3017 \pm 0.0013 \\
 $a_3/a_0$  & 0.2  &    &  0.2069 \pm 0.0025 \\
 \hline
\end{tabular}
\caption{Results of simulations. \label{tab:res_mc}}
\end{center}
  \end{table}

\section{Possible extensions}\label{sec:ext}
This subsection discusses some extensions which can be applied to the weighting/fitting method but in general 
not easily to the other methods.

If the polarisation vector points for example in an arbitrary unknown direction 
$\vec{P}=P(\cos(\varphi),\sin(\varphi))$ in the $x$-$y$ plane, 
the observed signal is
\begin{eqnarray}
  N(\varphi) &\propto& (1 + \epsilon_c \cos(\varphi) + \epsilon_s \sin(\varphi)) \, .
\end{eqnarray}
In this case, in the analysis one has to include also the sums
\[
  \sum_{\ua} \sin(\varphi_i)^n  \quad \mbox{and} \quad  \sum_{\da} \sin(\varphi_i)^n \quad \mbox{for}\quad  n=1,2 \, .
  \]
  This gives in total 10 equations for 10 unknowns.
  The unknowns are
\[
  (L^\ua \sigma_0 a_0), (L^\da \sigma_0 a_0), \frac{a_1}{a_0}, \frac{a_2}{a_0}, \frac{a_3}{a_0}, \frac{b_1}{a_0}, \frac{b_2}{a_0}, \frac{b_3}{a_0}, \epsilon_c  \, \, \mbox{and} \, \, \epsilon_s \, .
\]
Including also tensor polarisation for a spin 1 particle, 
the event distributions reads
\[
   N(\varphi) \propto (1 + \epsilon_c \cos(\varphi) + \epsilon_s \sin(\varphi) + \epsilon_{2c} \cos(2\varphi) + \epsilon_{2s} \sin(2\varphi)) \, .
\]
This problem can be solved by using the observables 
\[
 N,\, \sum_i \sin^n(\varphi_i) \, , \, \sum_i \cos^n(\varphi_i) \, ,   \quad \mbox{for}\quad  n=1,2,3,4 \, .
\]
for now in total three polarisation states. The number of equations increases to 27
for 19 parameters
\begin{eqnarray*}
&(L^\ua \sigma_0 a_0), (L^\da \sigma_0 a_0), (L^0 \sigma_0 a_0),& \\
&  \frac{a_1}{a_0},  \frac{a_2}{a_0}, \frac{a_3}{a_0}, \frac{a_4}{a_0}, \frac{a_5}{a_0}, \frac{a_6}{a_0}, &\\
&  \frac{b_1}{a_0}, \frac{b_2}{a_0}, \frac{b_3}{a_0}, \frac{b_4}{a_0}, \frac{b_5}{a_0}, \frac{b_6}{a_0}, &\\
&  \epsilon_c , \epsilon_s ,  \epsilon_{2c} \,\, \mbox{and} \, \, \epsilon_{2s} \, .& 
\end{eqnarray*}

Looking at eq.~\ref{eq:nup} to \ref{eq:nup2}, one observes that the parameter $a_3$ appears only once and even suppressed
with respect to $a_1$ by a factor 3. One could set $a_3$ to zero resulting in a fit with 6 equations for 5 unknowns,
which makes a $\chi^2$ test possible.
It is also possible to add a data set with unpolarized beam to the fit. This is for example useful
if the two polarisations $P^\ua$ and $P^\da$ are different and not known. 

It is interesting to note that the method introduced here, especially for the case were the number of equations exceeds the number of parameter is a special case of the ``Generalized Method of Moments'' (GMM) widely used in economics (e.g. see ref.~\cite{hansen,hansen_nobel}).

\section{Summary and conclusion}% Comparison of Methods

Two types of estimators to extract azimuthal asymmetries have been compared.
One is based on event counts and one on event weighting.
It was shown that estimators just using event counts do not use the full information contained in the data.
This is reflected in the fact that the figure of merit is smaller than in methods where events are weighted
with an appropriate weight.
The optimal weight for azimuthal asymmetries is $\cos(\varphi)$. It can also be shown that using this weight, 
the FOM is the
same as in a maximum likelihood method reaching the Cram\'er-Rao limit of the lowest possible statistical error.

Among the estimators using event weights the method introduced in this paper has the advantage
that no knowledge about the acceptance is required and no correction due to possible
difference in luminosity has to be applied. On the contrary, the method even provides
information on the azimuthal dependence of the acceptance.
The method is easily extendable to more observables.

%\begin{acknowledgements}
\section*{Acknowledgements}
The authors would like to thank M.~Hartmann for comments and discussions on the paper.
This work was triggered by discussions on polarimetry for a storage ring electric dipole moment (EDM) measurement pursued by the JEDI\footnote{
\href{http://collaborations.fz-juelich.de/ikp/jedi/}{http://collaborations.fz-juelich.de/ikp/jedi/}} collaboration and was supported by the ERC Advanced Grant (srEDM \#694340)
of the European Union.
%\end{acknowledgements}

\appendix

\section{Covariance matrix of observables}\label{app:cov}
The covariance matrix for the observables
\[
\vec{y}_{\mathrm obs} = \left( N^\ua, \sum_{\ua} \cos(\varphi_i),\sum_{\ua} \cos^2(\varphi_i),
   N^\da, \sum_{\da} \cos(\varphi_i),\sum_{\da} \cos^2(\varphi_i) \right)
   \]
   is
   \begin{eqnarray*}  
  C&=&
  \left(
  \begin{array}{cc}
    C_{\ua} &  0 \\
    0  &  C_{\da}
    \end{array}
  \right)
  \quad \, \quad \mbox{with}\\
%%%
  C_{\ua (\da)} &=&
  \left(
  \begin{array}{ccc}
    N^{\ua (\da)} & \sum_{\ua (\da)} \cos(\varphi_i) &\sum_{\ua (\da)} \cos^2(\varphi_i)\\
    \sum_{\ua (\da)}  \cos(\varphi_i) & \sum_{\ua (\da)} \cos^2(\varphi_i) &\sum_{\ua (\da)} \cos^3(\varphi_i)\\
    \sum_{\ua (\da)}  \cos^2(\varphi_i) & \sum_{\ua (\da)} \cos^3(\varphi_i) &\sum_{\ua (\da)} \cos^4(\varphi_i)\\    
    \end{array}
  \right) \, .
  \end{eqnarray*}
  A derivation of the correlation between sums over events for different weights used here can be found in ref.~\cite{Pretz:2011qb}
  (appendix A). 

\section{Figure of merit for cross ratio counting and weighting/fitting method}\label{app:fom_w}

\subsection{FOM of counting methods}\label{app:fom_cnt}
Assuming $P^\ua = -P^\da$ and $\avrg{\cos(\varphi)}_L = -\avrg{\cos(\varphi)}_R =:\avrg{\cos(\varphi)}$,
eq.~\ref{eq:A_cr} simplifies to
\[
   \hat{\epsilon} = \frac{1}{\avrg{\cos(\varphi)}} \, \frac{\sqrt{\delta}-1}{\sqrt{\delta}+1} \, .
\]
Applying standard error propagation, one finds
\[
  \sigma_{\epsilon}  = \frac{\dd \epsilon}{\dd \delta} \, \sigma_{\delta}
  \]
  with
\[
 \frac{\dd \epsilon}{\dd \delta} = \frac{1}{\avrg{\cos(\varphi)}}  \,   \frac{1}{(1+\sqrt{\delta})^2 \sqrt{\delta}}
 \]
 and
 \[
   \sigma_{\delta} = \sqrt{\frac{1}{N^\ua_L} + \frac{1}{N^\ua_R} + \frac{1}{N^\da_L} + \frac{1}{N^\da_R} } \quad  \delta \, .
 \]
 Using
 \[
 N^\ua_{L,R} = \frac{N_{\mathrm tot}}{2} \frac{1 \pm \avrg{\cos(\varphi)} \epsilon}{2}  \, , \quad \mbox{and} \, \,  N^\da_{R,L} = \frac{N_{\mathrm tot}}{2} \frac{1 \pm \avrg{\cos(\varphi)} \epsilon}{2}  \, ,
 \]
with $N_{\mathrm tot} = N^\ua_{L} + N^\ua_{R} + N^\da_{L} + N^\da_{R} $,
 one finds
 \[
\frac{ \sigma_{\delta}}{\delta}  = \frac{4}{\sqrt{N_{\mathrm tot}}} \, \frac{1}{\sqrt{1-\avrg{\cos(\varphi)}^2\epsilon^2}} \, .
 \]
 Using
 \[
  \sqrt{\delta} = \frac{1+\avrg{\cos(\varphi)} \epsilon}{1 - \avrg{\cos(\varphi)} \epsilon}
 \]
we finally arrive at
 \begin{eqnarray}
   \sigma_{\epsilon} &=& \frac{1}{\avrg{\cos(\varphi)}} \, \frac{1}{(1+\sqrt{\delta})^2 \sqrt{\delta}} \,  \frac{4}{\sqrt{N_{\mathrm tot}}} \, \frac{1}{\sqrt{1-\avrg{\cos(\varphi)}^2\epsilon^2}} \delta \\
                   &=&  \frac{1}{\avrg{\cos(\varphi)}} \,  \sqrt{\frac{1- \avrg{\cos(\varphi)}^2 \epsilon^2 }{N_{\mathrm tot}}} \,    \, .
   \end{eqnarray}
 The FOM is given by
 \[
  \mbox{FOM}_{\epsilon} = N_{\mathrm tot} \frac{\avrg{\cos(\varphi)}^2}{1-\avrg{\cos(\varphi)}^2 \epsilon^2}
 \]
 which agrees with eq.~\ref{eq:fom_cr_exact}.
For the estimators in eq.~\ref{eq:A_linear} the FOM is obtained by a similar procedure.
 
\subsection{FOM of weighting methods}
Defining the luminosity factor $\ell_0 = \mathcal{L} \sigma_0 a_0$,
equations~\ref{eq:chi2} can be linearized around $\ell^\ua=\ell_0$,
$\ell^\da=\ell_0$, $\epsilon=\epsilon_0$ for $a_1=a_2=a_3=0$.
Resulting in a system of linear equations
\[
  y_{\mathrm model} = A x_{\mathrm para} + y_0
\]
with
\begin{eqnarray*}
x_{\mathrm para} &=& (\Delta \ell^\ua, \Delta \ell^\da,\Delta \epsilon)^T \, , \quad 
y_0 = \ell_0 \left(1, \frac{1}{2} \epsilon_0, \frac{1}{2},1, \frac{1}{2}  \epsilon_0, \frac{1}{2} \right)^T \\
A&=& \,
\left(
\begin{array}{ccc}
 1 & 0 & 0 \\
 \epsilon_0 c_2 & 0 & \ell_0 c_2 \\
 c_2 & 0 & 0 \\
 0 & 1 & 0 \\
 0 & -\epsilon_0 c_2 & -\ell_0 c_2 \\
 0 & c_2 & 0 \\
\end{array}
\right) \, \quad \mbox{and} \, \\
C &=&
\ell_0 \, 
\left(
\begin{array}{cccccc}
 1 & c_2 {\epsilon_0} & c_2 & 0 & 0 & 0 \\
c_2 \epsilon_0 & c_2  & c_4 \epsilon_0  & 0 & 0 & 0 \\
 c_2 & c_4 \epsilon_0 & c_4 & 0 & 0 & 0 \\
 0 & 0 & 0 & 1 & - c_2 \epsilon_0 & c_2  \\
 0 & 0 & 0 & -c_2 \epsilon_0 & c_2 & -c_4 \epsilon_0 \\
 0 & 0 & 0 & c_2 & -c_4\epsilon_0 & c_4 \\
\end{array}
\right)   \, ,
\end{eqnarray*}
with $c_n = \avrg{\cos(\varphi)^n}$.
The covariance matrix $C$ is the same as in app.~\ref{app:cov} except that we used here
the expectation values instead of sum over events to arrive at an analytic expression.
The covariance matrix for the parameters $(\Delta \ell^\ua, \Delta \ell^\da,\Delta \epsilon)$, which is identical
to the covariance matrix for the parameters $(\ell^\ua, \ell^\da, \epsilon)$ since they just differ by a constant vector,
is given by:
\[
C_{\mathrm para} = (A^T \, C^{-1} \, A)^{-1} =
\left(
\begin{array}{ccc}
 \ell_0 & 0 & 0 \\
 0 & \ell_0 & 0 \\  
 0 & 0 & \frac{c_2 - c_4 \epsilon_0^2}{2 c_2^2 \ell_0} \\ 
\end{array}
\right) \, .
\]
For the FOM of $\epsilon_0$, replacing $\ell_0$ by $N_{\mathrm tot}/2$, one thus finds
\[
 \mbox{FOM}_\epsilon = N_{\mathrm tot} \frac{\avrg{\cos^2(\varphi)}^2} {\avrg{\cos^2(\varphi)} - \avrg{\cos^4(\varphi)} \epsilon_0^2} \, ,
\]
which agrees with eq.~\ref{eq:fom_w_exact}.

%\section{Analytic expressions for analyzing powers}\label{app:analytic_expr}
%
%The analytic expression for $A$ resulting from eq.~\ref{eq:chi2} is 
%\footnotesize
%\begin{equation}
%
%\hat{A} =\frac{  P^\da (c_1^\da c_1^\ua  - c_2^\ua N^\da) + P^\ua (- c_1^\da c_1^\ua + c_2^\da N^\ua ) - 
%   \sqrt{(-c_1^\da c_1^\ua P^\da + c_2^\ua N^\da P^\da + c_1^\da c_1^\ua P^\ua - 
%      c_2^\da N^\ua P^\ua)^2 - 
%    4 (c_1^\ua N^\da - c_1^\da N^\ua) (-c_1^\ua c_2^\da P^\da P^\ua + 
%       c_1^\da c_2^\ua P^\da P^\ua)}}{2 P^\da P^\ua c_1^\da c_2^\ua (1  -\frac{c_1^\ua c_2^\da}{c_1^\da c_2^\ua} )}
%%
%
%  
%\hat{A} =\frac{ (c_1^\da c_1^\ua P^\da - c_2^\ua N^\da P^\da - c_1^\da c_1^\ua P^\ua + c_2^\da N^\ua P^\ua - 
%   \sqrt{(-c_1^\da c_1^\ua P^\da + c_2^\ua N^\da P^\da + c_1^\da c_1^\ua P^\ua - 
%      c_2^\da N^\ua P^\ua)^2 - 
%    4 (c_1^\ua N^\da - c_1^\da N^\ua) (-c_1^\ua c_2^\da P^\da P^\ua + 
%       c_1^\da c_2^\ua P^\da P^\ua)})}{2 (-c_1^\ua c_2^\da P^\da P^\ua + 
%     c_1^\da c_2^\ua P^\da P^\ua)}
%\end{equation}
%\normalsize

\bibliography{statistics.bib,literature_dis.bib}

\bibliographystyle{ieeetr}

\end{document}